\documentclass[aps,prl,twocolumn,showpacs,floatfix,preprintnumbers,amsmath,amssymb]{revtex4-1}
\usepackage{amsmath}
\usepackage{amssymb}
\usepackage{graphicx}
\usepackage{epsfig}
\usepackage{dcolumn}
\usepackage{bm}
\usepackage{array}
\usepackage{xcolor}
\usepackage{multirow}

\begin{document}
\title{Symmetry deduction from spectral fluctuations in complex quantum systems}
\author{S. Harshini Tekur}
\email{E-mail: harshini.t@gmail.com}
\affiliation{Max-Planck-Institut f\"ur Physik komplexer Systeme,
N\"othnitzer Stra\ss e 38, 01187 Dresden, Germany}
\author{M. S. Santhanam}
\email{E-mail: santh@iiserpune.ac.in}
\affiliation{Indian Institute of Science Education and Research, 
Dr. Homi Bhabha Road, Pune 411 008, India}

\begin{abstract}

The spectral fluctuations of complex quantum systems, in appropriate limit, are known
to be consistent with that obtained from random matrices. However, this relation
between the spectral fluctuations of physical systems and random matrices is valid
only if the spectra are desymmetrized. This implies that the fluctuation properties of
the spectra are affected by the discrete symmetries of the system. In this work, it 
is shown that in the chaotic limit the fluctuation characteristics and symmetry 
structure for any arbitrary sequence of measured or computed levels can be 
inferred from its higher-order spectral statistics
without desymmetrization. In particular, we consider a spectrum composed of $k>0$ independent
level sequences with each sequence having the same level density. The $k$-th order spacing ratio distribution of such
a composite spectrum is identical to its nearest neighbor counterpart with modified Dyson index $k$.
This is demonstrated for the spectra obtained from random matrices, quantum billiards, spin chains
and experimentally measured nuclear resonances with disparate symmetry features.

\end{abstract}

\maketitle

Spectral fluctuations in complex quantum systems are analyzed
using the theoretical framework of random matrix theory (RMT) in many
areas of physics \cite{guhr_physrep, porter_book, haake, akemann2011oxford, forrester2010}. These include few-body systems studied
in quantum chaos \cite{stockmann} to interacting many-body systems in condensed matter physics \cite{rmt-cmp},
nuclear \cite{kota_physrep} and atomic physics \cite{rmt-atphys}. These fluctuations carry signatures
of the distinct phases observed in the physical systems, {\it viz.,} integrable or 
chaotic limit of the underlying classical system \cite{reichl}, metallic or insulating phase \cite{rmt-mit},
localized or thermal phase of many-body systems \cite{rmt-mbl}, low-lying shell model or mixing 
regime of nuclear spectra \cite{rmt-np, rev_brody}. Indeed, the level spacing distribution of the {\sl desymmetrized} 
eigenlevels is a popular diagnostic tool to discriminate the between phases of physical systems in many
areas of physics, and remarkably even outside of physics \cite{akemann2011oxford, outphys1, outphys2, outphys3, outphys4}.

Beginning with the Wigner surmise \cite{mehta2004} in the context of nuclear spectra,
the present consensus is that the spectral fluctuations of complex quantum 
systems, in suitable limit, display level repulsion consistent with that of an
appropriately chosen ensemble of random matrices. For the special case of
quantum chaotic systems, the Bohigas-Giannoni-Schmidt (BGS) conjecture encapsulates 
this connection between the spectra from physical systems and random matrices \cite{bgs}. This has 
been amply verified in experiments \cite{bgs_exp},
simulations \cite{bgs_sim} and derives some theoretical support based on semiclassical techniques \cite{semiclass}.

\begin{figure*}
\includegraphics*[width=3.3in]{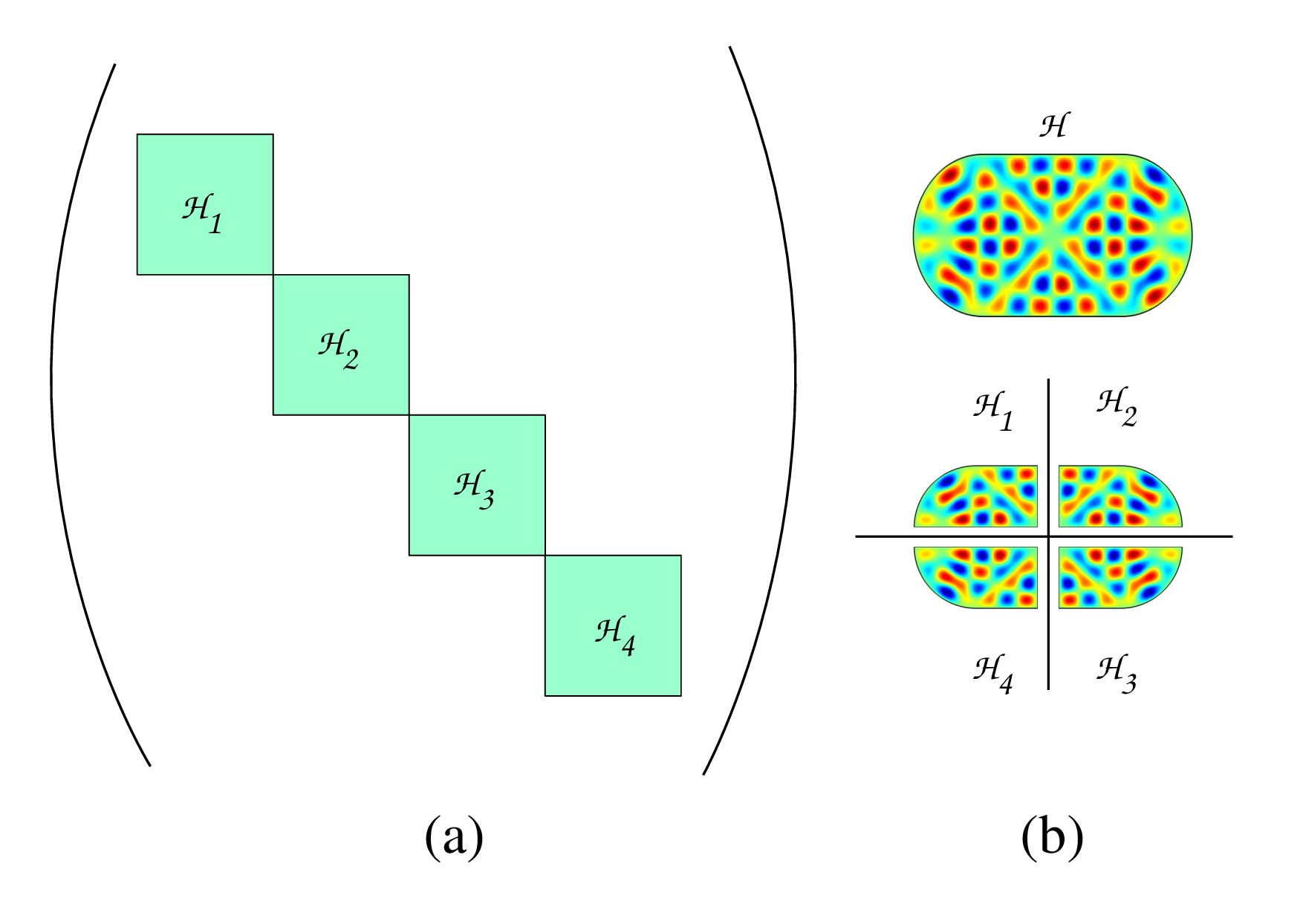}
\includegraphics*[width=3.3in]{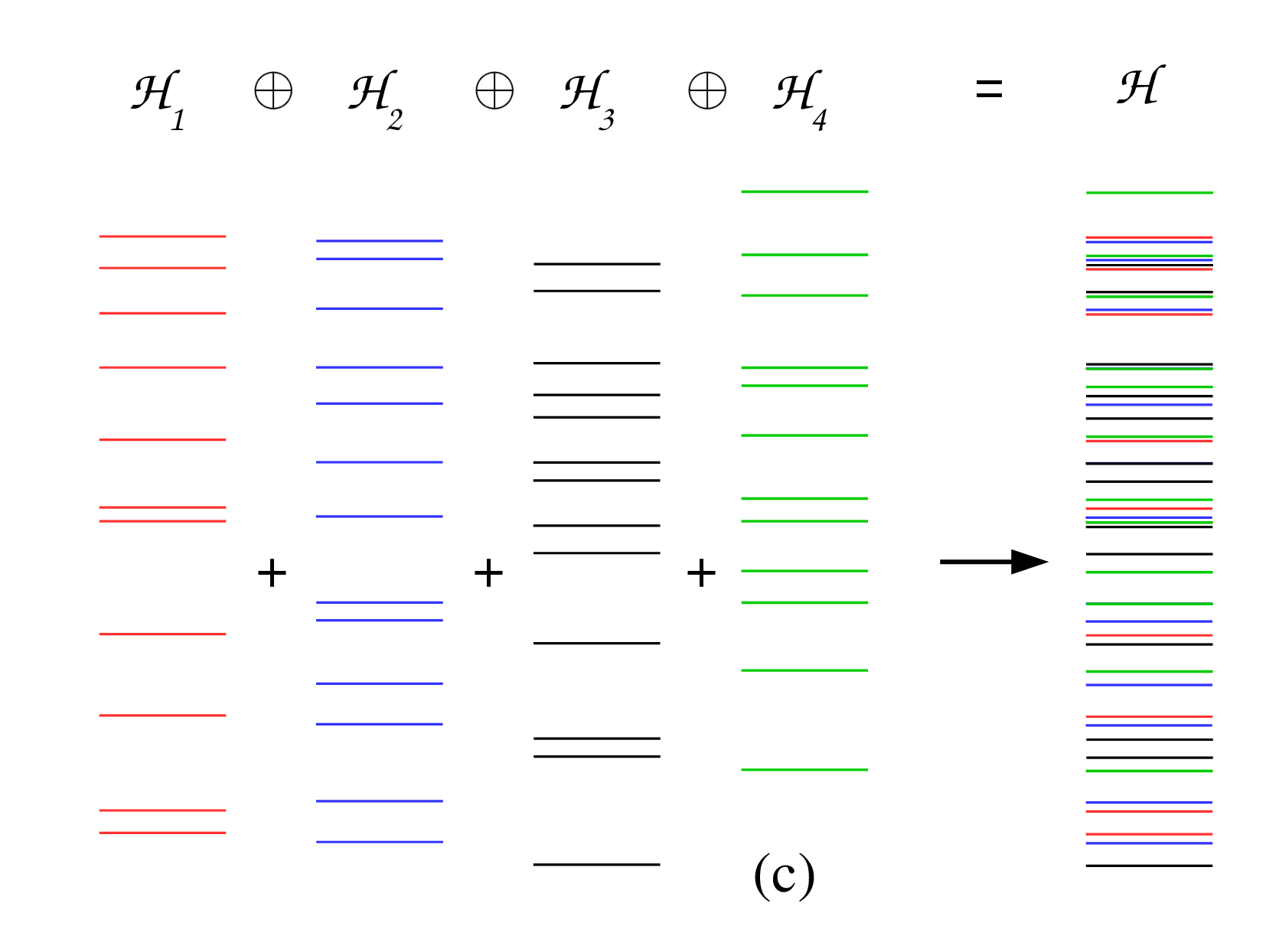}
\caption{(a) A schematic of the Hamiltonian matrix for stadium billiards with each block characterized 
by a good quantum number. (b) The potential for the stadium billiard with one of 
its eigenstates superposed on it. The bottom part shows a desymmetrized version of the same eigenstate indicating
one possible relation to the blocks in the Hamiltonian in (a).
(c) Schematic of the eigenlevels arising from each symmetry block, $\mathcal{H}_1$ (red coloured levels) to 
$\mathcal{H}_4$ (green).  The levels of $\mathcal{H}$ in the last column are a superposition of all these 
desymmetrized levels.}
\label{schematic}
\end{figure*}

Discrete symmetries of the system, {\it i.e.}, invariance of the potential under parity, reflection,
rotation, are crucial in realizing this connection between spectral fluctuations and
dynamical phases. In the presence of symmetries, the
Hilbert space of the system splits into invariant subspaces or the Hamiltonian matrix $\mathcal{H}$
becomes block diagonal, {\it i.e.}, $\mathcal{H} = \mathcal{H}_1 \oplus \mathcal{H}_2 \oplus \dots \mathcal{H}_m$, with each 
block $\mathcal{H}_i, i=1,2 ...m$ characterized by good quantum numbers corresponding to 
the respective symmetries \cite{stockmann}. This is schematically shown for chaotic billiards, with $m=4$ symmetry sectors,
in Fig. \ref{schematic}.
To compute any measure of spectral fluctuation, all the discrete levels
must be drawn from the same subspace (shown as blocks in Fig. \ref{schematic}(a)).
If symmetries are ignored and levels from 
different blocks are superposed, as depicted in Fig. \ref{schematic}(c), the genuine correlation
between levels (that should have produced level repulsion) is masked by near-degeneracies
resulting in level clustering. This effect becomes even more dominant as the number of superposed spectra
$m$ increases. This is misleading since level clustering is also a
spectral signature of integrable systems \cite{berry-tabor}.

This implies that the level correlations are sensitive to the presence or absence
of symmetries. It is then reasonable to expect that fluctuations of composite spectra, superposed
from many independent blocks, contain information about the entire system's symmetry structure.
However, any measure based on the nearest neighbor (NN) fluctuations, such as the popular 
NN level spacing distribution, will always tend to the Poissonion limit
(level clustering) due to the superposition of non-interacting blocks \cite{berry-robnik}.
In this work, rigorous numerical evidence is presented to show that the higher-order level spacing ratio
not only identifies the true fluctuation character, viz, level clustering or repulsion, 
but also allows us to deduce quantitative information about the symmetry structure of the composite Hamiltonian matrix $\mathcal{H}$.


This result obviates the need for symmetry decomposition of quantum systems, allows
for the analysis of any arbitrary sequence of experimentally observed levels with unknown symmetry structure, 
and is also of considerable interest in RMT \cite{forrester}.
Let $G$ be a random matrix
such that $G = G_1 \oplus G_2 \oplus \dots G_m$, a superposition of
$m$ blocks each of which is a Gaussian random matrix with identical level density. Given an arbitrary sequence of
eigenvalues of $G$, the fluctuation properties and the block
structure of $G$ can be inferred from its higher-order fluctuation statistics.
The proposed method is straightforward, involving only the calculation
of spacing ratios. This is in contrast to the cumbersome
methods proposed earlier based on two-level cluster function and requiring regression to 
deduce $m$ from any composite spectrum \cite{mehta2004, porter_book, symded,guhr_physrep},
all of which require unfolding as the first step.

Consider a sequence of eigenvalues $E_i, i=1,2,\dots N$ of a quantum  operator or a random matrix.
Spectral fluctuations are relatively easier to analyze using spacing ratios defined
as $r_i=\frac{E_{i+2}-E_{i+1}}{E_{i+1}-E_i}, i=1,2 \cdots N-2$ \cite{huse}, as spacing ratios are independent of the local
density of states and hence do not require spectral unfolding. For random matrix ensembles with Dyson
index $\beta=1, 2$ and 4, corresponding respectively
to the Gaussian orthogonal, unitary and symplectic ensembles, the distribution of
spacing ratios is given by \cite{atas}
\begin{equation}
P(r,\beta)=C_\beta \frac{(r+r^2)^\beta}{(1+r+r^2)^{1+3\beta/2}},
\label{rd1}
\end{equation}
where $C_\beta$ is a constant, as listed in Ref. \cite{atas}.
These RMT models are applicable to Hamiltonians with time-reversal invariance (TRI) ($\beta=1$),
without TRI ($\beta=2$) and TRI with spin-1/2 interactions ($\beta=4$).
The main focus of this paper is on RMT models applicable to Hamiltonians with TRI ($\beta=1$).
For integrable systems, the ratio distribution becomes $P_P(r)=1/(1+r)^2$.
Recently, an expression for nearest neighbor spacing ratio distribution has been obtained
taking into account the spectral transition from integrable to chaotic limits, and also crossovers
from one symmetry class to another \cite{corps}.

\begin{figure}[t]
\includegraphics[width=3.3in]{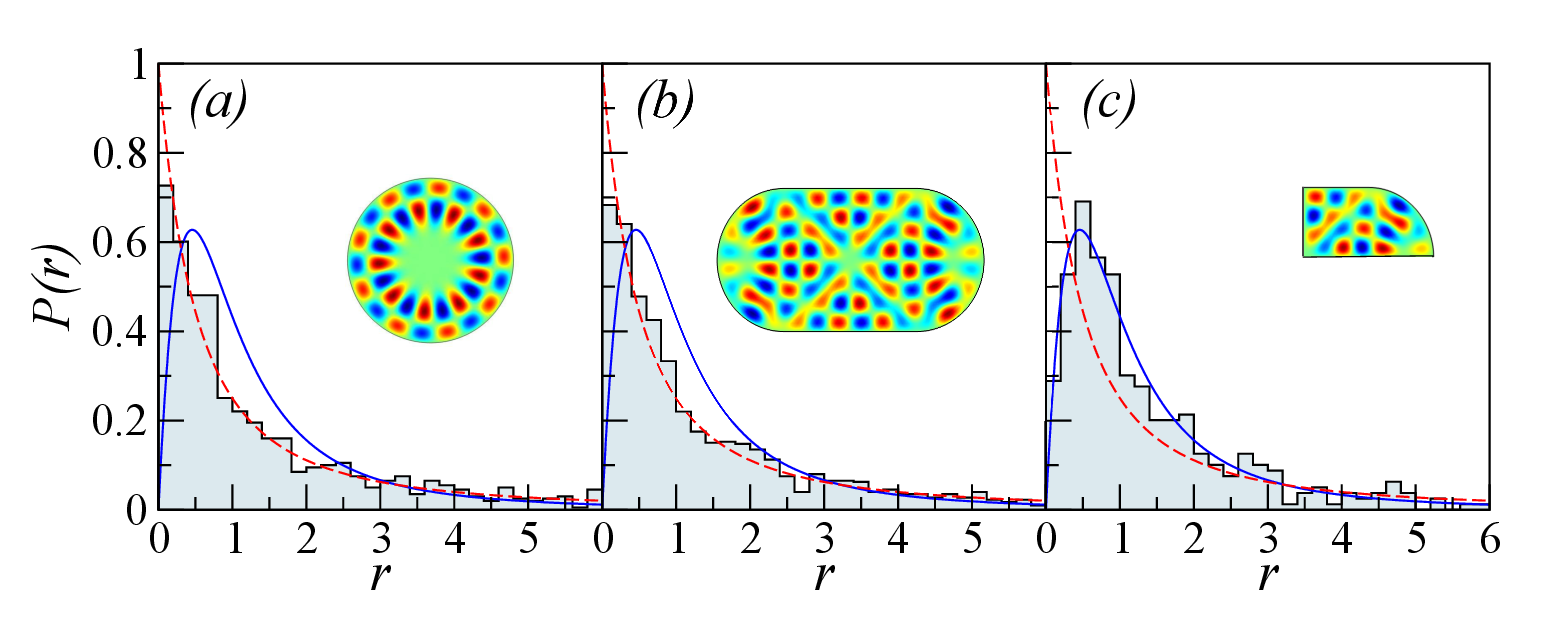}
\caption{Distribution of the NN spacing ratios (histograms) $P(r)$ for the (a) circular,
(b) stadium, and (c) desymmetrized stadium billiards. The dashed (red) line is $P_P(r)$
and the solid (blue) curve is the Wigner surmise for ratios. The inset shows
the shape of billiards and a typical eigenfunction superposed on it to emphasize its
symmetry structure.}
\label{fig1}
\end{figure}

As motivation, in Fig. \ref{fig1}, the numerically computed distribution of NN 
spacing ratios $P(r)$ is shown for circular (integrable) \cite{circular} and 
stadium (chaotic) \cite{stadbill} billiards. The integrable billiards (Fig. \ref{fig1}(a)) expectedly
agrees with $P_P(r)$. Note that stadium billiard has $C_{2v}$ point group symmetry with four 
irreducible representations (irreps). If the spectra from each irrep is analyzed 
separately, by BGS conjecture, an agreement with $P(r,1)$
of GOE is observed (Fig. \ref{fig1}(c)). However, in Fig. \ref{fig1}(b), the spectra from all the 
irreps is superposed, and hence the ratio distribution is closer to $P_P(r)$ with pronounced
deviation from $P(r,1)$.
In quantum systems with chaotic limit, as demonstrated below, the true character of their spectral fluctuations and
the number $m$ of independent spectra superposed can all be inferred using only the higher-order spacing ratio (HOSR)
distributions without {\it apriori} knowledge of its symmetry structure.

To this end, we consider non-overlapping $k$-th order spacing ratio, defined as
\begin{align}
r_i^{(k)} = \frac{s_{i+k}^{(k)}}{s_{i}^{(k)}} = \frac{E_{i+2k}-E_{i+k}}{E_{i+k}-E_i}, ~~~~
i,k=1,2,3,\dots .
\label{hosr}
\end{align}
In what follows, spectra from $m$ independent blocks are superposed, and its distribution 
of $k$-th order spacing ratios is denoted by $P^k(r,\beta,m)$. We consider only $\beta=1$.
For the special case involving NN ratios, we denote $P^1(r,\beta,1) = P(r,\beta)$.
The motivation for considering higher-order fluctuation statistics arises from 
a seminal result conjectured in Ref. \cite{dyson} and proved by Gunson \cite{gunson} for
the case of circular ensembles of RMT. If two independent spectra from the circular
orthogonal ensemble (COE) are superposed, upon integrating out every alternate eigenvalue, 
the joint probability distribution of the remaining eigenvalues follow circular unitary ensemble 
(CUE) statistics. In terms of higher-order measures, this result states that the second order
statistics of two superposed COE spectra converges to NN statistics of CUE.
This is reflected in the distribution of spacings and spacing ratios as well.
In the limit of large matrix dimensions, this result holds for Gaussian ensembles too
yielding $P^2(r,1,2)=P(r,2)$ for two superposed spectra.
If the order of each of the $m$ GOE matrices is the same, then this may be generalized for the 
superposition of $m$ GOE spectra as
\begin{equation}
P^k(r,1,m)=P(r,\beta'), ~\mbox{where}~ \beta'=m=k,
\label{maineqn}
\end{equation}
implying that its $k$-th order spacing ratio distribution converges to NN 
statistics $P(r,\beta')$ with $\beta'=k$. Equation \ref{maineqn} is the main result of the paper, and 
is well supported by numerical experiments involving a small number of mixed symmetries (up to $m=5$).
In contrast, irrespective of how many {\it uncorrelated} spectra are superposed corresponding 
to integrable systems, the $k$-th order spacing ratio distribution can be obtained 
(details in supplementary information \cite{supp}) as
\begin{equation}
P_P^k(r)=\frac{(2k-1)!}{[(k-1)!]^2}\frac{r^{k-1}}{(1+r)^{2k}}.
\label{hopd}
\end{equation}
For $k=1$, this reduces to $\frac{1}{(1+r)^2}$, the correct limit for the NN spacing
ratio for uncorrelated spectra.
We note that Eq. \ref{maineqn} is reminiscent of a scaling relation reported recently in Ref. \cite{tekur}.

For the superposition of $m=2$ to $5$ independent GOE spectra, validity of Eq. \ref{maineqn} is verified
in Fig. \ref{fig2} . In this figure, an excellent agreement is observed between histograms 
obtained from the computed eigenvalues of GOE matrices and the solid line representing 
$P(r,\beta'=k)$. For uncorrelated eigenvalues, a similar agreement with Eq. \ref{hopd} is observed.
In order to independently obtain a best quantitative estimate for $\beta'$ in
Eq. \ref{maineqn} for a given superposition of $m$ spectra, we compute
\begin{align}
D(\beta')=\sum_i ~ \bigl| I^m_{obs}(r_i,1,m) - I(r_i,\beta') \bigl|.
\end{align}
In this, $I^m_{obs}(r,1,m)$ and $I(r,\beta')$ represents the cumulative distribution
functions corresponding respectively to the observed histogram $P^k(r,1,m)$ and 
the postulated function $P(r,\beta')$.
If the minima of $D(\beta')$ occurs at, say, $\beta'=\beta_0$,
then $\beta_0$ is the best estimate consistent with the observed data.
As seen in the insets of Fig. \ref{fig2}, the minima in  $D(\beta')$ coincides
with the value of $m$, the number of superposed spectra. This is further corroborated by
the Kolmogorov-Smirnov (KS) test \cite{kstest} at a significance level of 0.05 for each case.
The $p$-value, as anticipated, is maximum at the same value of $m$.

\begin{figure}[t]
\includegraphics[width=3.3in]{Fig3.eps}
\caption{Distribution of $k$-th order spacing ratios (histograms) for a superposition of
$m$ GOE spectra, each obtained from matrix of order $N=40000$, shown for
$m=2$ to $5$. The solid curve is $P(r,\beta')$, with $\beta'=k$. The insets
show $D(\beta)$ whose minimum correctly coincides with the expected value of $m$.
The $p$-values from KS test are also given.}
\label{fig2}
\end{figure}

A complete picture is revealed in Fig. \ref{rmtfit} for a superposition of $m=4$ independent
GOE spectra, where the computed histogram for the $k$-th order ratio is shown for
$k=2$ to $7$. Based on Eq. \ref{maineqn}, we expect it to be consistent with $P(r,\beta'=4)$.
For each $k$, $P^k(r,1,4)$ is matched against the corresponding $P(r,\beta')$, and
$D(\beta')$ is calculated. Both visually and quantitatively (the minima of 
$D(\beta')$ in Fig. \ref{rmtfit}(e)), best agreement is observed for $k=4$, verifying 
the main result in Eq. \ref{maineqn}.
Significantly, for the superposed spectra, Eqs. \ref{maineqn}-\ref{hopd} can be used to infer 
the correct nature of spectral fluctuations (level repulsion or clustering) and also to
determine the number of superposed independent blocks for a random matrix or the number 
of diagonal blocks in the Hamiltonian matrix of a complex quantum system, if the system is chaotic.
In what follows, the result in Eqs. \ref{maineqn}-\ref{hopd} will be applied to 
chaotic systems possessing different symmetries, notably billiards and spin chains, and
most importantly to the experimentally measured data of nuclear resonances.

\begin{figure}[]
\includegraphics*[width=3.0in]{Fig4a.eps}
\includegraphics*[width=3.0in]{Fig4b.eps}
\caption{(a-f) Computed $k$-th order spacing ratio distribution (histogram) for superposed spectra
from four GOE matrices of order $N=40000$.
Note that the best agreement with $P(r,\beta'=k)$ (blue line) obtained only for $\beta'=k=4$, 
and is confirmed by the $p$-value from the KS test, displaying maxima at $\beta'=4$. (g) $D(\beta')$ vs. $\beta'$ has minima
at $\beta'=4$. Both $D(\beta')$ and $p$ indicate the validity of Eq. \ref{maineqn}.}
\label{rmtfit}
\end{figure}

\begin{figure}[t!]
\includegraphics[width=3.3in]{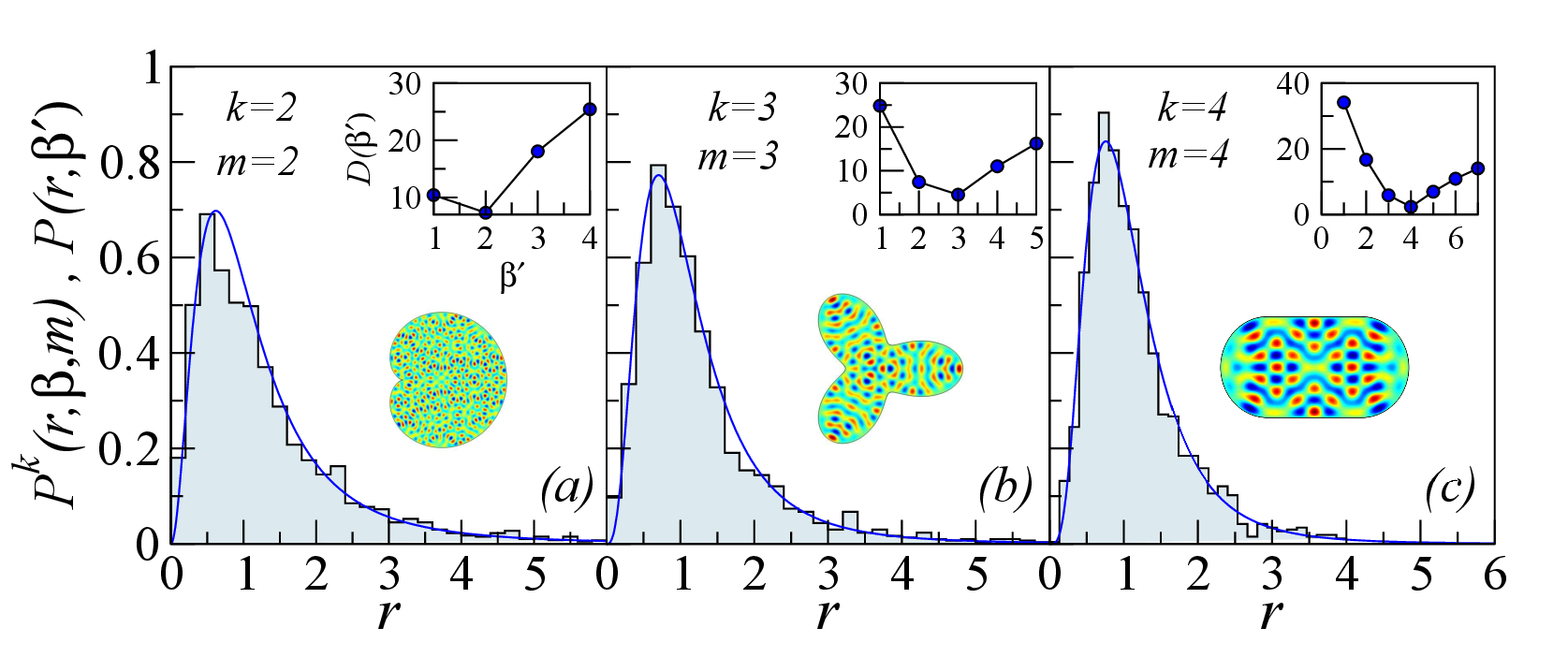}
\caption{HOSR distribution (histogram) for the billiards family in which
spectra from (a) $k=2$, (b) $k=3$ and (c) $k=4$ irreps are superposed. The higher-order
distributions are consistent with $P(r,\beta')$, and $\beta'=k$ as dictated by Eq. \ref{maineqn}.
The insets display $D(\beta')$ and its minimum indicates the correct number of
irreps in the system. Also shown is the shape of billiards with an arbitrarily
chosen chaotic eigenstate to highlight its symmetry.}
\label{billiards}
\end{figure}

First we consider quantum billiards, in which a free particle is confined in a 
cavity defined by a variety of boundaries \cite{robnik}, whose eigenspectrum is
obtained by solving the Helmholtz equation with Dirichlet boundary conditions. 
They are popular models in Hamiltonian chaos and mesoscopic physics \cite{berry_bill}
and have experimentally-realized variants \cite{bill_exp}. Modifying the boundary or shape
of the billiard changes its symmetry and also drives it from 
integrability to chaos. 
For a billiard whose boundary is parameterized by $r(\phi)=r_0(1+\epsilon \cos\phi)$, 
as $\epsilon$ varies from
0 to 1, the system transitions from integrable to chaotic dynamics. For $\epsilon=0$, 
a circular billiard shown in Fig. \ref{fig1}(a) is obtained. This is an integrable system and its
higher-order spacings are in agreement with Eq. \ref{hopd} (See Ref. \cite{supp}). For $\epsilon=1$,
the cardioid billiard is obtained \cite{cardioid},
possessing two irreps due to reflection symmetry about the horizontal axis. Thus, eigenlevels 
obtained disregarding symmetry would correspond to a superposition of two GOE spectra. 
As anticipated by Eq. \ref{maineqn}, its second order distribution $P^2(r,1,2)$ is
consistent with $P(r,2)$ (Fig. \ref{billiards}(a)).  A billiard with three irreps, similar
in shape to one that has been experimentally realized \cite{dietz_3foldbill}, is 
obtained by parameterizing its boundary as $r(\phi)=r_0(1+0.3\cos(3\phi))$.
This model, with symmetries ignored and after removing degeneracies arising from the
two-dimensional irreps,
corresponds to a superposition of three chaotic spectra and the best match for $P^3(r,1,3)$ is provided
by $P(r,3)$ (Fig. \ref{billiards}(b)). A chaotic
billiard with four irreps is the well-studied Bunimovich stadium billiard \cite{stadbill1} shown in Fig. \ref{billiards}(c). 
This has reflection symmetry about $x$ and $y$ axes
and, in accordance with Eq. \ref{maineqn}, $P^k(r,1,4)$ displays the best correspondence with $P(r,\beta')$ 
for $k=\beta'=4$ (Fig. \ref{billiards}(c)). For all of these cases, insets in Fig. \ref{billiards} show that
the minima of $D(\beta')$ coincides with $\beta'=k$, the number of irreps.

Next, a spin-$1/2$ chain with the Hamiltonian \cite{santos}
 \begin{align}
H&=\sum_{i=1}^{L-1}[J_{xy}(S_i^xS_{i+1}^x+S_i^yS_{i+1}^y)+J_zS_i^zS_{i+1}^z] \nonumber \\
   &+\eta \sum_{i=1}^{L-2}[J'_{xy}(S_i^xS_{i+2}^x+S_i^yS_{i+2}^y)+J'_zS_i^zS_{i+2}^z] \label{schain1}
\end{align}
is considered, where $L$ is the number of sites, $J_{xy}$ and $J_z$ are the NN coupling 
strengths in three directions (coupling along $x$ and $y$ being the same), and $J'_{xy}$ 
and $J'_z$ are the next NN coupling strengths. This system is integrable for $\eta=0$ 
(as shown in Fig. 1 in Ref. \cite{supp}), and chaotic for
$\eta \gtrsim 0.2$. The total spin in the $z$-direction, $S_z$, is conserved and the Hamiltonian is block
diagonal in $S_z$ basis, each block corresponding to a given value of $S_z$. 
However, other symmetries that exist in this system would erroneously lead to fluctuation statistics appearing 
to be integrable in this subspace (not shown here).
For odd number of sites ($L_{odd}$), on computing 
the HOSRs and comparing with corresponding $P(r,\beta')$, $k=\beta'=2$ has the 
best match (Fig. \ref{spinchain}(a)). But for even number of sites ($L_{even}$), HOSRs correspond 
to $k=\beta'=4$ (Fig. \ref{spinchain}(b)). This is because for $L_{odd}$ or $L_{even}$, the 
parity operator (with eigenvalues $\pm1$) commutes with $H$, leading to two invariant subspaces 
in a given $S_z$ block. For $L_{even}$, an additional rotational symmetry exists (with 
eigenvalues $\pm 1$) for the corresponding operator giving rise  to four irreps. The other parameters
used in Figs. \ref{spinchain}(a,b) are $J_{xy}=J'_{xy}=1.0$, $J_z= J'_z=0.5$,
with $L_{even}=14$ and $L_{odd}=15$. For these systems, KS test results given in \cite{supp}
provide yet another verification of the scaling relation.

\begin{figure}[t!]
\includegraphics[width=3.3in]{spin_chains.eps}
\caption{HOSR distribution computed for the spin-1/2 chain
Hamiltonian in Eq. \ref{schain1}, with (a) odd number of sites with two irreps and
(b) even number of sites with four irreps. The insets show $D(r,\beta')$ and its
minima identifies the number of irreps.}
\label{spinchain}
\end{figure}

Even for systems whose Hamiltonian is not well-defined or unknown as in the
case of complex nuclei, experimentally observed nuclear resonance data can be analyzed to 
characterize its fluctuation statistics and find its number of irreps. It is assumed that the system being
observed is in the putative regime in which RMT results can be applied. We consider a sequence 
of experimentally observed neutron resonances for $Ta^{181}$ nucleus \cite{Ta} whose NN spacing distribution, 
discussed in Ref. \cite{rev_brody}, does not match the Wigner 
surmise. On calculating HOSR distributions, remarkably, Eq. \ref{maineqn} holds 
good for $k=2$, further confirmed by the minima of $D(\beta')$ for $\beta'=2$ in 
Fig. \ref{Ta181}, and the corresponding KS test results. This indicates the presence of two independent
symmetry sectors, and is indeed the case, as confirmed in Refs. \cite{Ta, rev_brody}. This measured sequence consists 
of a superposition of levels having angular momentum $J=3$ and $4$, and when symmetry decomposed, are in
broad agreement with Wigner surmise. Clearly, for an arbitrary sequence of 
measured levels, if random matrix description is valid, HOSRs based on Eq. \ref{maineqn} can unambiguously 
identify the true fluctuation character and the number of symmetry sectors.

In experiments, often measurement errors lead to missing levels \cite{missing_levels} and hence
incorrect identification of the fluctuation character and number of irreps. 
The robustness of Eq. \ref{maineqn} to missing levels in a
superposition of GOE spectra was tested in two ways; (a) by randomly deleting levels, (b) by preferentially
deleting one of a pair that is nearly degenerate.
Upon computing $D(\beta')$ in each case (details in \cite{supp}), the scaling in Eq. \ref{maineqn}
holds good even if $20-30\%$ (40\%) of the levels are removed through random deletions (deletion of
near-degenerate levels). This is because higher-order fluctuations are unaffectected by, and hence
largely insensitive to, randomly missing levels. This virtue is inherent to this method and has
practical significance for analyzing experimental data.


\begin{figure}[t]
\centerline{\includegraphics*[width=3.0in]{Ta181a.eps}}
\centerline{\includegraphics*[width=3.0in]{Ta181b.eps}}
\caption{(a-d) The $k$-th order spacing ratio distribution (histogram) for experimentally 
observed nuclear resonances for Tantalum (Ta$^{181}$), showing the best correspondence for $k=2$.
The broken line is $P(r,\beta'=k)$. The $p$-value from KS test shows a maximum at $\beta'=2$. 
(e) $D(\beta')$ shows minima at $\beta'=2$, reinforcing the validity of Eq. \ref{maineqn}.}
\label{Ta181}
\end{figure}

To summarize, quantum systems must be symmetry decomposed to reveal its true
spectral fluctuation characteristics. This also implies that the fluctuations carry symmetry information,
though extracting it unambiguously from NN fluctuation statistics is non-trivial. As demonstrated in this work,
the higher order spacing ratio distributions can reveal, apart from the fluctuation
characteristics, quantitative information about symmetry structure. For a superposition
of $k$ independent spectra (with identical level densities) drawn from an ensemble of RMT, the 
central result (Eq. \ref{maineqn}) relates the $k$-th order spacing ratio distribution for random matrices 
with Dyson index $\beta=1$ to the corresponding nearest neighbor statistics with $\beta'=k$. For quantum systems
in the classically chaotic limit and in the regime of applicability of Wigner-Dyson ensembles
of RMT, this elegant relation determines the number of irreps (or diagonal blocks) present
in a Hamiltonian matrix. This is exploited to analyze any arbitrary 
sequence of experimentally measured or computed levels, even if the system's Hamiltonian
and symmetry structure are unknown. This technique requires neither unfolding nor 
free-parameter estimation, nor computation of cumbersome correlation or power spectral
functions and hence straightforward to implement.
Further, for uncorrelated eigenvalues, the HOSR distribution has been derived
and can be used as a test of integrability. These results are
demonstrated using disparate physical systems like quantum billiards, spin chains
and experimentally measured nuclear resonances.
It must be remarked that just as Bohigas-Giannoni-Schmidt conjecture \cite{bgs} is valid for
quantum systems in their chaotic limit, the results presented here too are strictly valid in the 
same regime in which RMT is applicable. In principle, this approach may be extended to 
weakly chaotic or mixed systems (outside of RMT regime) by considering a broader
class of higher-order spacing ratios and these results will be reported elsewhere.

\onecolumngrid
\newpage

\begin{center}
 
{\Large \bf \underline{Supplemental Material}}
\end{center}

\section{Higher order distribution of spacing ratios for a sequence of uncorrelated eigenvalues (eigenvalues of integrable quantum systems)}

\subsection{Analytical expression}

For a given sequence of uncorrelated eigenvalues, $E_1\leq E_2\leq \cdots E_N$, the spacings between nearest neighbours is defined as 
$s_i=E_{i+1}-E_i, i=1,2, \cdots N$. The distribution of these spacings is of the form $P(s)=e^{-s}$, and hence distributions of spacings
and spacing ratios for integrable quantum systems are termed Poissonian. \\

The ratios of nearest neighbour spacings for these systems are defined as $r_i=s_{i+1}/s_i, i=1,2, \cdots N$, and the distribution
of these ratios is of the form $P(r)=1/(1+r)^2$\cite{ABGR2013}.\\

Ratios of higher order spacings may be defined as
\begin{align}
 r_i^{(k)} = \frac{s_{i+k}^{(k)}}{s_{i}^{(k)}} = \frac{E_{i+2k}-E_{i+k}}{E_{i+k}-E_i}, ~~~~~~~
i,k=1,2,3,\dots .
\label{hosr}
\end{align}

To obtain a form for the distribution of $r^{(k)}$, the higher order spacings may be expressed in terms of nearest neighbour spacings as
\begin{align}
 s_{i}^{(k)}&=E_{i+k}-E_i \\ \nonumber
            &=E_{i+k}-E_{i+k-1}+E_{i+k-1}-E_{i+k-2}+ \cdots +E_i \\ \nonumber
            &= s_{k}+ \cdots +s_{i+1}+s_i.  
\end{align}

Then the distribution of $s_i^{(k)}$ may be calculated as the distribution of a sum of $k$ random variables $s_i$, each of which is
distributed as $P(s)=e^{-s}$. For simplicity, $s_i^{(k)}$ is denoted as $z$ below. The distribution of $z$ is given by 
\begin{equation}
P(z)= \frac{e^{-z}z^{k-1}}{(k-1)!}
\label{sumrv}
\end{equation}

Then the distribution of higher order spacing ratios is simply the distribution of the quotient of two random variables, each of which is distributed
as Eq. \ref{sumrv}. This distribution may be calculated as
\begin{equation}
 P_P^{(k)}(r)=\int |z|P(rz)P(z)dz
\end{equation}

Substituting for $P(z)$ and $P(rz)$ from  Eq. \ref{sumrv}, 
\begin{align}
  P_P^{(k)}(r)&=\int_0^\infty |z| \frac{e^{-rz}(rz)^{k-1}}{(k-1)!}  \frac{e^{-z}z^{k-1}}{(k-1)!} dz \nonumber \\ 
              &= \frac{r^{k-1}}{(k-1)!^2}\int_0^\infty z^{2k-1} e^{-z(r+1)} dz.
\end{align}

This can be evaluated in terms of the incomplete gamma function $\Gamma(x)$ as
\begin{align}
 P_P^{(k)}(r)&=\frac{\Gamma(2k)}{(k-1)!^2}\frac{r^{k-1}}{(1+r)^{2k}} \nonumber \\
             &= \frac{(2k-1)!}{\big((k-1)!\big)^2}\frac{r^{k-1}}{(1+r)^{2k}}.
\end{align}

For $k=1$, it reduces to the familiar form 
\begin{equation*}
 \frac{1}{(1+r)^2}.
\end{equation*}

For $k=2$,
\begin{equation}
P_P^{(2)}(r)=\frac{6r}{(1+r)^4},
\end{equation}
 for $k=3$,
\begin{equation}
 P_P^{(3)}(r)=\frac{30r^2}{(1+r)^6},
\end{equation}
and for $k=4$,
\begin{equation}
 P_P^{(4)}(r)=\frac{140r^3}{(1+r)^8}.
\end{equation}

\subsection{Comparison of analytical form of $P_P^{(k)}(r)$ with results from physical systems}

\vspace{0.2cm}

\begin{figure}[h]
\begin{center}
 \includegraphics[scale=0.5]{integrable_new.eps}
\end{center}
\caption{Higher order spacing ratio distributions for $k=2$ to $4$, for uncorrelated eigenvalues (upper panel, indigo), circular billiards (lower panel, red)
and integrable spin chain obtained by setting $\eta=0$ in Eq. 6 of the main paper (lower panel, black). The corresponding analytical result (Eq. 4 in the
main paper) is also shown in all cases (upper and lower panels, broken blue curve).}
\end{figure}

\section{Results of Kolmogorov-Smirnov Test for billiards and spin chains of different irreps}

The $p$-values of the Kolmogorov-Smirnov (KS) test for the billiards corresponding to Fig. 5 of 
the main text, as well as the spin chains corresponding to Fig. 6 of the main text, are given below.
This further confirms the validity of the main result of the paper. In general, at 5\% significance level,
the KS-test could not reject the hypothesis that the data are consistent with the distribution in Eq. 3, which
is the main result of this paper.

\begin{table}[t!]
\begin{tabular}{|c|c|c|}
\toprule
System			             & k                & p \\ \hline
\multirow{3}{*}{\textit{Billiards}}  & 2              & 0.735               \\ [3pt]
                                     & 3              & 0.671               \\ [3pt]
                                     & 4              & 0.706               \\ [3pt] \hline
\multirow{2}{*}{\textit{Spin chain}} & 2              & 0.730               \\ [3pt]
                                     & 4              & 0.929               \\ \hline
\end{tabular}
\caption{The {\it p}-values ({\it p}) for the KS test
at a significance level of 0.05, for the billiards in Fig. 5 and the spin chains in Fig. 6 of the main text corresponding to
the superposition of $k$ irreducible representations.}
\label{tab:tab1}
\end{table}

\section{Effect of missing levels}

The effect of missing levels in a given sequence of superposed spectra is studied in two ways. First,
by randomly deleting a fixed percentage of levels, and then calculating higher order spacing ratios, 
from a superpostion of GOE spectra of dimension $N=40000$. Second, in experiments it is often difficult to
resolve near-degeneracies in a spectrum. Again, this leads to the problem of missing levels, which is
simulated by deleting a fixed percentage of one of two nearly-degenerate eigenvalues randomly in a superposition
of GOE spectra, as before. In each case, $D(\beta')$ is calculated, and the value of $\beta'$ corresponding
to the minima of $D(\beta')$ corresponds to the best fit.

\vspace{1cm}

\begin{figure}[h]
 \begin{center}
  \includegraphics[scale=0.5]{missing_levels.eps}
 \end{center}
\caption{$\beta'$ (for which $D(\beta')$ is minimum) as a function of percentage of missing levels (diamonds) and 
a percentage of near-degeneracies (circles), obtained by evaluating the second (fourth) order spacing ratio distribution
for a superposition of two(four) GOE spectra, plotted in red(blue).} 
\label{miss}
\end{figure}

Fig. \ref{miss} shows the value of $\beta'$ (evaluated in steps of 0.1) plotted against the percentage of missing levels (diamonds)
as well as nearly degenerate levels (circles), when $P^k(r,1,m)$ is evaluated for a superposition of $m$ GOE spectra,
where $m=2$ (blue) and $m=4$ (red). According to Eq. 4 of the main paper, namely,
$P^k(r,1,m)=P(r,\beta'), ~\mbox{where}~ \beta'=m=k$, the expected value of $\beta'$ is 2 (for $k=2$) and 4 (for $k=4$) respectively.
For the case where a percentage of levels are randomly deleted (full lines with diamonds), it may be observed that assuming
even a $10\%$ fluctuation in the numerical evaluation of $\beta'$, a significant deviation from the
predicted $\beta'$ occurs only when about $20\%$ of the levels are missing. A similar behavior was seen for spin chains
with 2 and 4 irreps as well (not shown). However, the value of $\beta'$ remains robust against randomly deleting upto $45\%$ of levels
which differ in magnitude with one of their nearest neighbours by a factor of $10^{-4}$. This is shown by the dashed lies with circles.

\end{document}